\documentstyle[preprint,aps]{revtex}
\begin{document}
%
\title{
Spin conductance, dynamic spin stiffness and spin diffusion in itinerant magnets
}
\author{Peter Kopietz\cite{address}}
\address{
Institut f\"{u}r Theoretische Physik der Universit\"{a}t G\"{o}ttingen,\\
Bunsenstrasse 9, D-37073 G\"{o}ttingen, Germany}
\date{August 25, 1997}
\maketitle
\begin{abstract}
We discuss analogies between the charge- and spin response
functions of itinerant magnets. 
We show that the spin-analog of the charge stiffness
is {\it{not}} given by the usual spin stiffness
$\rho_s$, but by the {\it{dynamic spin stiffness}} $D_s$, which is
obtained from the dynamic {\it{spin conductance}} $G_s ( \omega )$ in
the limit of vanishing frequency $\omega$.
The  low-frequency behavior of $G_s ( \omega)$ is used to define   
ideal spin conductors, normal
spin conductors, and spin insulators.
Assuming diffusive spin dynamics, we show that the
spin diffusion coefficient
is proportional to
$\lim_{\omega \rightarrow 0} {\rm Re} {G}_s ( \omega )$.
We exploit this fact to develop a 
new extrapolation scheme for
the spin diffusion coefficient in the paramagnetic
phase of the  Hubbard model.

\end{abstract}
\narrowtext
%
%
\section{Introduction}
\setcounter{section}{1}
\label{sec:intro}
In a classic paper on the {\it{Theory of the Insulating State}}
W. Kohn\cite{Kohn64} pointed out that 
the behavior of the charge stiffness $D_c$ can be
used to distinguish the conducting from the insulating
state of interacting electrons.
Physically $D_c$ can be identified with the weight of the Drude
peak of the frequency-dependent conductivity $\sigma ( \omega)$.
A system with a finite value of $D_c$ 
deserves to be called an
{\it{ideal
conductor}}, because then  $\sigma (
\omega)$
diverges as $  D_c / (i \omega )$ for $\omega \rightarrow 0$.
In contrast, a {\it{normal conductor}} 
does not have a Drude peak ($D_c = 0$), and the conductivity
$\sigma ( \omega )$ approaches a non-zero value for vanishing
frequency. 
Finally, the {\it{insulating}} state  can be characterized 
by $D_c = 0$ and
$\sigma ( \omega = 0) = 0$.
Note that $D_c$ is defined in terms of the conductivity,
a dynamic quantity. 
Nevertheless, $D_c$ can be obtained
without explicitly calculating $\sigma ( \omega)$:
as shown by Kohn\cite{Kohn64},
the second derivative of the free energy
with respect to a fictitious vector potential
(which is equivalent to a twist in the boundary conditions
on the many-body wave-function)
is proportional to the charge stiffness.
More recently, the arguments of Kohn\cite{Kohn64}
have been sharpened by Shastry and Sutherland\cite{Shastry90},
who calculated $D_c$ exactly for Heisenberg-Ising and Hubbard rings
using the ansatz of Bethe.

Another important observable,  which is closely related to
$D_c$, is the superfluid stiffness
 $\rho_c$ (the motivation for our slightly unconventional notation
will become obvious below).   
Whereas $D_c$ characterizes the normal conducting
properties of the system, 
a finite value of $\rho_c$ indicates long-range
superconducting correlations. 
$D_c$ and $\rho_c$ can both be obtained  
as different limits of
the wave-vector  and frequency-dependent
current response function 
$K_{\alpha \beta} ( {\bf{q}} , \omega )$, which is
given by the Kubo formula\cite{Scalapino92}.
We shall briefly summarize the relevant definitions in
Sec.\ref{sec:charge}.

While $D_c$ and $\rho_c$  in Hubbard and related
models have recently
been studied  numerically by several groups\cite{Scalapino92,Castella95,Zotos96},
the analogous quantities $D_s$ and $\rho_s$ that characterize the
{\it{spin}} dynamics
have not received much attention. In fact,
some authors seem not to be fully aware of 
the physically different meaning of the
spin analog $D_s$ of the charge stiffness on the one hand,
and the spin analog $\rho_s$ 
of the superfluid stiffness on the other hand.
The quantity
$\rho_s$ is commonly called the {\it{spin stiffness}},
and is related to the change in energy due to a small static twist
in the directions of the spins at the boundaries of the
system\cite{Fisher73}. A finite value of $\rho_s$
in the thermodynamic limit
is a manifestation of (quasi)-long-range  magnetic order.
Obviously,
the spin stiffness $\rho_s$ and the superfluid stiffness $\rho_c$
are analogous quantities: both measure the
degree of off-diagonal long-range order in the system.
But what is the spin analog of the charge stiffness, and
what is its physical meaning?
In this paper we shall not only answer this question, but also
discuss the more general concept of the
spin conductance $G_s ( \omega)$.
We show that the limiting behavior of $G_s ( \omega)$
for vanishing frequency $\omega$ can be used 
to define {\it{ideal spin conductors}}, {\it{normal spin conductors}}, and
{\it{spin insulators}}, in complete analogy with the charge transport.
Normal spin conductors are of particular
interest, because systems with diffusive spin dynamics
fall into this category. 
In Sec.\ref{sec:diffusion} we shall 
use the concept of  the spin conductance to
develop a new extrapolation scheme for calculating
the spin diffusion coefficient of these systems.
For localized spin models (such as the Heisenberg model)
we have discussed the spin conductance
and its usefulness for the calculation
of the spin diffusion coefficient in Ref.\cite{Kopietz93}.

For definiteness we shall consider the repulsive Hubbard model,
although our considerations are easily generalized for other
itinerant magnets.
The Hamiltonian is given by
$\hat{H} = \hat{T} + \hat{V}$, with
 \begin{eqnarray}
 {\hat{T}} & = &
 -  t \sum_{\bf{r}} 
 \sum_{\alpha = 1}^{d}
 [ c^{\dagger}_{\bf{r}} c_{{\bf{r}} + {\bf{a}}_{\alpha} }  + {\rm h.c.} ]
 \; \; \; ,
 \label{eq:Tdef}
 \\
 \hat{V} & = & U \sum_{\bf{r}} \left[
 c^{\dagger}_{{\bf{r}} \uparrow} c_{{\bf{r}} \uparrow} - \frac{1}{2} 
 \right]
 \left[ c^{\dagger}_{{\bf{r}} \downarrow} c_{{\bf{r}} \downarrow}
 - \frac{1}{2} \right]
 \label{eq:Vdef}
 \; \; \; ,
 \end{eqnarray}
where 
${\bf{r}}$ labels the $N$ sites of a $d$-dimensional
hyper-cubic lattice, and $ {\bf{a}}_{\alpha}$
are the primitive vectors with length $a$ in direction
$\alpha = 1 , \ldots , d$.
The operators
$c^{\dagger}_{ {\bf{r}} \sigma }$, $\sigma = \uparrow , \downarrow$ 
create spin-$\sigma$ electrons at lattice site ${\bf{r}}$, and
$c^{\dagger}_{\bf{r}} = [ c^{\dagger }_{{\bf{r}} \uparrow} , 
c^{\dagger}_{{\bf{r}} \downarrow} ]$
are two-component operators.

\section{Charge response: charge and superfluid stiffness}
\label{sec:charge}

Let us first recall the Kubo formula for the
conductivity, which measures the linear charge response
to an electromagnetic field. 
The usual frequency-dependent conductivity 
can be written as 
 $\sigma_{\alpha \beta } ( \omega )
 = 
 \frac{e^{2}}{h} a^{2-d}  
 G_{\alpha \beta} ( \omega )$,
where the dimensionless conductance\cite{footnote1}
 $G_{\alpha \beta} ( \omega )$ can be obtained from the
current response function 
 $K_{\alpha \beta} ( {\bf{q}} , \omega )$,
 \begin{equation}
 G_{\alpha \beta} ( \omega  )
  =  \lim_{ {\bf{q}} \rightarrow 0} \frac{   K_{\alpha \beta} ( {\bf{q}} , \omega +
  i 0^{+}) }{ i ( 
 \hbar \omega  + i 0^{+})} 
 \; \; \; .
 \label{eq:conductance}
 \end{equation}
The current response function has two contributions,
 \begin{equation}
 K_{\alpha \beta} ( {\bf{q}} , \omega + i 0^{+})
  = 
  - \delta_{\alpha \beta} D^{\rm dia}
 +  P_{\alpha \beta} ( {\bf{q}} , \omega + i 0^{+}) 
 \label{eq:Kubocharge}
 \; \; \; ,
 \end{equation}
where the diamagnetic part $D^{\rm dia}$ 
is proportional to the
thermal expectation value of the kinetic energy operator, 
 \begin{equation}
 D^{\rm dia} =  \frac{ 1}{ d N } \langle  - \hat{T}  \rangle 
 \equiv
 \frac{ 1}{ d N } \sum_{n} p_n \langle n |   - \hat{T}  | n \rangle 
 \label{eq:Ddia}
 \; \; \;  .
 \end{equation}
and the paramagnetic contribution is for general complex frequency $z$ given by
 \begin{eqnarray}
  P_{\alpha \beta} ( {\bf{q}} , z ) & = &  
 \frac{1}{N}
 \sum_{n,m} p_n
 \left[ 
 \frac{
  \langle n| {\hat{J}}_{\alpha} ( {\bf{q}} ) | m  \rangle
  \langle m | {\hat{J}}_{\beta} ( - {\bf{q}} ) | n \rangle }
 {E_{m} - E_{n} - \hbar z }
 \nonumber
 \right.
 \\
 & & 
 \left.
 \hspace{10mm} +
 \frac{
  \langle n| {\hat{J}}_{\beta} ( - {\bf{q}} ) | m  \rangle
  \langle m | {\hat{J}}_{\alpha} ( {\bf{q}} ) | n \rangle }
 {E_{m} - E_{n} + \hbar z  }
 \right]
 \; .
 \label{eq:rhopara}
 \end{eqnarray}
Here $| n \rangle$ denotes a complete set of exact eigenstates of
$\hat{H}$, and $p_n = ( \sum_m e^{- E_m / T } )^{-1}  e^{- E_n/T}$ are the
thermal occupation probabilities of states with  energies $E_n$,
where $T$ is the temperature (measured in units of energy).
The current operators are
 \begin{equation}
 {\hat{J}}_{ \alpha } ({\bf{q}}) =   \frac{t}{ 2 i} \sum_{\bf{r}}
 e^{- i {\bf{q}} \cdot {\bf{r}} }
 \left[ {c}^{\dagger}_{\bf{r}} 
 ( {c}_{{\bf{r}} + {\bf{a}}_{\alpha} }  -
  {c}_{{\bf{r}} - {\bf{a}}_{\alpha} } ) 
  - \mbox{h.c.}
 \right]
 \label{eq:Jxdef}
 \; \; \; .
 \end{equation}
The charge  stiffness tensor $[ D_c ]_{\alpha \beta}$ and the 
superfluid stiffness tensor $[ \rho_c ]_{\alpha \beta}$
can be defined via the following limiting procedures,
 \begin{eqnarray}
 {[ D_c ]}_{\alpha \beta }&  = & - \lim_{\omega \rightarrow 0}
 \left[ \lim_{\bf{q} \rightarrow 0} K_{\alpha \beta } ( {\bf{q}} , \omega + i 0^{+} ) 
 \right]
 \label{eq:Dcdef}
 \; \; \; ,
 \\
 {[ \rho_c ]}_{ \alpha \beta } & =  &
 - \lim_{ {\bf{q}} \rightarrow 0}
 \left[ \lim_{\omega \rightarrow 0} 
  K_{\alpha \beta} ( {\bf{q}} , \omega + i 0^{+} ) 
 \right]
 \label{eq:rhocdef}
 \; \; \; .
 \end{eqnarray}
For completeness, let us also introduce the 
${\bf{q}}$-dependent superfluid stiffness\cite{Tinkham75},
\begin{equation}
 [ \rho_c ( {\bf{q}} ) ]_{\alpha \beta } =
 - \lim_{\omega \rightarrow 0} K_{\alpha \beta } ( {\bf{q}} , \omega + i 0^{+} ) 
 \label{eq:rhockdef}
 \;  ,
 \end{equation}
which probes the response to  
time-independent transverse electromagnetic fields (see below).
Eqs.(\ref{eq:conductance}), (\ref{eq:Dcdef})-(\ref{eq:rhockdef}) 
relate physical quantities characterizing the
charge dynamics to the appropriate limits of the linear response function
$K_{\alpha \beta} ( {\bf{q}} , \omega)$.
For convenience, we have
chosen a gauge where the scalar potential is set equal to zero, so that
the electric field is represented by a vector potential,
${\bf{E}} (t) = - c^{-1} \partial {\bf{A}} (t) / \partial t$. 
This gauge, which is used very often in the literature,
has the advantage that the current response can be expressed 
entirely in terms of the current-current correlation function.
Of course, the physical current response is gauge invariant,
see Ref.\cite{Martin68}.

\begin{sloppypar}
Quite generally, at long wavelengths ($| {\bf{q}} | a \ll 1$)
the current response tensor $K_{\alpha \beta} ( {\bf{q}} , \omega )$
can be decomposed into a longitudinal  part
($K_{\|}$) and a transverse part ($K_{\bot}$), i.e. 
\end{sloppypar}
 \begin{equation}
 K_{\alpha \beta} ( {\bf{q}} , \omega )
 = {\hat{{q}}}_{\alpha} {\hat{{q}}}_{\beta} K_{\|} ( {\bf{q}} , \omega )
 + ( \delta_{\alpha \beta} -
  {\hat{{q}}}_{\alpha} {\hat{{q}}}_{\beta} )
 K_{\bot} ( {\bf{q}} , \omega )
 \; ,
 \label{eq:Kdecomp}
 \end{equation}
where $\hat{q}_{\alpha} = \hat{\bf{a}}_{\alpha} \cdot \hat{\bf{q}}$,
with $\hat{\bf{a}}_{\alpha} = {\bf{a}}_{\alpha} / a$ and
${\hat{\bf{q}}} = {\bf{q}} / q$.
The corresponding decomposition for the
tensor $[ \rho_c ( {\bf{q}} ) ]_{\alpha \beta}$ 
contains only a transverse component,
 \begin{equation}
[ \rho_c ( {\bf{q}} ) ]_{\alpha \beta} =
  ( \delta_{\alpha \beta} -
  {\hat{{q}}}_{\alpha} {\hat{{q}}}_{\beta} ) \rho_c ( {\bf{q}} )
\; .
\label{eq:rhocdecomp}
\end{equation}
The longitudinal part vanishes. Physically, this is due
to the fact that  a static
longitudinal vector potential cannot induce any current\cite{Pines89}.
If we set ${\bf{q}} = 0$ in Eq.(\ref{eq:Kdecomp}) we have
 $K_{\bot} ( 0 , \omega )
=
 K_{\|} ( 0 , \omega )$, because for a spatially uniform field
the longitudinal and transverse response are identical.
Thus, the conductance tensor
$G_{\alpha \beta} ( \omega )$ in Eq.(\ref{eq:conductance})
and the charge stiffness tensor $[D_c]_{\alpha \beta}$ in Eq.(\ref{eq:Dcdef}) are
proportional to the unit matrix (for a system with
cubic symmetry).
The usual charge stiffness $D_c$ and superfluid stiffness
$\rho_c$ can be identified with  eigenvalues of the corresponding
ternsors (\ref{eq:Dcdef},\ref{eq:rhocdef}), i.e.
$[ D_c ]_{\alpha \beta }  = \delta_{\alpha \beta } D_c$,
and $ [ \rho_c  ]_{\alpha \beta}  =  
  ( \delta_{\alpha \beta} -
  {\hat{{q}}}_{\alpha} {\hat{{q}}}_{\beta} ) \rho_c$. 
Note that with our normalization both quantities have units of energy.
$\rho_c$ is proportional to the
density of the superconducting
electrons.
The finite value of $\rho_c$ in a superconductor 
is closely related to the screening of the magnetic field,
i.e. the Meissner effect. Note that in a normal metal
$\rho_c = 0$, which is a consequence of the fact that
in the normal metallic state static magnetic fields are not screened.
Finally, let us point out that
the physical meaning of the
different order of limits in Eqs.(\ref{eq:Dcdef})
and (\ref{eq:rhocdef}) is easy to
understand from the Maxwell equation
$ c {\bf{q}} \times {\bf{E}} ( {\bf{q}} , \omega ) =
\omega {\bf{B}} ( {\bf{q}} , \omega ) $:
If we first take the limit $\omega \rightarrow 0$, 
the electric field ${\bf{E}}$ vanishes while
the magnetic field ${\bf{B}}$ can remain finite -- in this way we probe the Meissner effect.
On the other hand, if we first let ${\bf{q}} \rightarrow 0$, we are
left with an electric field.

\section{Spin response: Dynamic and static spin stiffness}
\label{sec:spin}

The generalization of the above definitions for the spin 
degrees of freedom is straightforward.
Denoting by $\sigma^{i}$, $i= x,y,z$, the Pauli matrices,
in $d$ dimensions we may define $3 \times d$ spin-current operators,
 \begin{eqnarray}
 {\hat{J}}_{ \alpha }^{i} ({\bf{q}}) & = &  \frac{t}{2 i} \sum_{\bf{r}}
 e^{-i  {\bf{q}} \cdot {\bf{r}} }
 \left[ {c}^{\dagger}_{\bf{r}} 
 \frac{\sigma^{i}}{2}
 ( {c}_{{\bf{r}} + {\bf{a}}_{\alpha} }   -
  {c}_{{\bf{r}} - {\bf{a}}_{\alpha} }  ) 
  - \mbox{h.c.}
 \right]
 \; ,
 \nonumber
 \\
 & & 
 \hspace{20mm}
 i = x,y,z
\; \; \; , \; \;  \;
\alpha = 1 , \ldots , d  
 \label{eq:Jxdefspin}
 \; \; \; .
 \end{eqnarray}
In complete analogy with Eq.(\ref{eq:Kubocharge})
we define the retarded spin-current response function
 \begin{equation}
 {K}_{\alpha \beta}^{ ij} ( {\bf{q}} , \omega + i 0^{+})
 = 
  - \delta_{\alpha \beta} \delta_{ij} D^{\rm dia}
 + {P}_{\alpha \beta }^{ ij} ( {\bf{q}} , \omega + i 0^{+})
 \label{eq:Kubospin}
 \; \; \; ,
 \end{equation}
where the diamagnetic contribution $D^{\rm dia}$ is
given in Eq.(\ref{eq:Ddia}), and the paramagnetic
term 
 ${P}_{\alpha \beta }^{ ij} ( {\bf{q}} , z )$
is simply obtained from Eq.(\ref{eq:rhopara}) by replacing
$\hat{J}_{\alpha} \rightarrow \hat{J}_{\alpha}^{i}$ 
and
$\hat{J}_{\beta} \rightarrow \hat{J}_{\beta}^{j}$.
As discussed by Chandra, Coleman, and Larkin\cite{Chandra90},
the spin-current response function ${K}_{\alpha \beta}^{ij} ( {\bf{q}} , \omega )$
gives the spin response
to a fictitious vector potential 
$\delta{{{A}}}^{i}_{\beta} ( {\bf{q}} , \omega )$, which
describes a space- and time-dependent modulation in the 
local spin-density. 
The proper definition of the dynamic- and static
spin stiffness tensor is now evident,
 \begin{eqnarray}
 [ {D}_s ]^{ij}_{\alpha \beta } & = &
 - \lim_{\omega \rightarrow 0}
 \left[ \lim_{ {\bf{q}} \rightarrow 0}
 {K}_{\alpha \beta}^{ ij} ( {\bf{q}} , \omega + i 0^{+})
 \right]
 \label{eq:Dsdef}
 \; \; \;  ,
 \\
 {[{\rho}_s ]}^{ij}_{\alpha \beta } & = &
 - \lim_{ {\bf{q}} \rightarrow 0}
 \left[
 \lim_{\omega \rightarrow 0}
 {K}_{\alpha \beta}^{ ij} ( {\bf{q}} , \omega + i 0^{+})
 \right]
 \label{eq:rhosdef}
 \; \; \; .
 \end{eqnarray}
Furthermore, in analogy with 
the ${\bf{q}}$-dependent superfluid stiffness introduced in
Eq.(\ref{eq:rhockdef}), let us define the
${\bf{q}}$-dependent spin stiffness tensor\cite{Kopietz91}
 \begin{equation}
 {[{\rho}_{s} 
  ( {\bf{q}} ) ]}_{\alpha \beta}^{ij}  =
 - \lim_{\omega \rightarrow 0} {K}_{\alpha \beta }^{ ij  } ( {\bf{q}} , \omega + i 0^{+} ) 
 \label{eq:spinstiffk}
 \; .
 \end{equation}
Finally, in analogy with Eq.(\ref{eq:conductance})
we introduce the dimensionless spin conductance
 \begin{equation}
 {[ {G}_s ( \omega )]}_{\alpha \beta }^{ ij}
  = 
 \lim_{ {\bf{q}} \rightarrow 0 } 
 \frac{ {K}_{\alpha  \beta }^{  ij} ( {\bf{q}} , \omega + i 0^{+})}{ i ( 
 \hbar \omega  + i 0^{+})} 
 \; \; \; .
 \label{eq:spinconductivity}
 \end{equation}
Because 
the operators
${\hat{J}}_{ \alpha }^{i} (0)$ 
define the ferromagnetic
spin-currents, the limit ${\bf{q}} \rightarrow 0$
in Eqs.(\ref{eq:Dsdef}), (\ref{eq:rhosdef}) and (\ref{eq:spinconductivity})
implies that we are looking at {\it{ferromagnetic}} correlations.
In the case of {\it{antiferromagnetism}}
we simply should consider the limit ${\bf{q}} \rightarrow {\bf{\Pi}} $ instead,
where ${\bf{\Pi}} = [ \pi / a  , \ldots , \pi /a ]$ is the
antiferromagnetic ordering wave-vector.
A summary of the analogous quantities characterizing
the charge- and the spin dynamics is given in
Table \ref{tab:analogy}.
We would like to emphasize that 
the spin analog of the Drude weight $D_c$
is given by the dynamic spin stiffness
${D}_s$, and {\it{not}} by the static spin stiffness 
${\rho}_s$. 
It seems that the dynamic spin stiffness has not been
discussed in the literature on the Hubbard model.
Following the terminology used for the
charge dynamics, a system with $D_s > 0$ can be called 
an {\it{ideal spin conductor}}. 
For $D_s = 0$ and $G_s ( 0 ) \neq 0$
the system is a {\it{normal spin conductor}}, and
the {\it{spin insulator}} can be characterized
by $D_s = G_s (  0) = 0$.

\section{Spin diffusion}
\label{sec:diffusion}

\subsection{Spin diffusion coefficient
and Thouless number}

To see the connection between the spin conductance and spin diffusion,
consider the dynamic structure factor
for the spin degrees of freedom,
 \begin{eqnarray}
 {S}^{ij} ( {\bf{q}} , \omega ) & = &  \frac{ 2 \pi \hbar }{N} \sum_{n,m}
 p_n \delta ( E_m - E_n - \hbar \omega ) 
 \nonumber
 \\
 & \times &
 \langle n | \hat{{S}}^{i}_{\bf{q}} | m \rangle
 \langle m | \hat{{S}}^{j}_{- \bf{q}} | n \rangle
 \;  .
 \label{eq:dynstruc}
 \end{eqnarray}
Here the Fourier components of the spin operators are
 $\hat{S}^{i}_{\bf{q}} =  \sum_{\bf{r}} e^{- i   {\bf{q}} \cdot {\bf{r}}  }
 c^{\dagger}_{\bf{r}} \frac{\sigma^i}{2} c_{\bf{r}}$.
General hydrodynamic arguments\cite{Forster75} show that the
diffusive spin dynamics manifests itself 
in the following long-wavelength and low-energy
form of the dynamic structure factor, 
 \begin{equation}
 {S}^{ij} ( {\bf{q}} , \omega ) = 
 2 \delta^{ij} \chi   \frac{\hbar \omega}{ 1 - e^{- \hbar \omega / T } }
 \frac{   {\cal{D}} {\bf{q}}^2 }{  \omega^2 + (  {\cal{D}} {\bf{q}}^2 )^2 }
 \; \; \; ,
 \label{eq:spindifdyn}
 \end{equation}
where we have assumed cubic symmetry and spin-rotational invariance.
Here ${\cal{D}}$ is the spin diffusion coefficient, and
the static susceptibility $\chi$ is 
 \begin{equation}
 \chi  =  \frac{1}{T}  \lim_{ {\bf{q}} \rightarrow 0}  \int_{ - \infty}^{\infty} 
 \frac{d \omega}{2 \pi} {S}^{ii} ( {\bf{q}} , \omega )
  =  \frac{1}{T} \sum_{\bf{r}} 
 \langle \hat{{S}}^{i}_0  \hat{{S}}^{i}_{\bf{r}} \rangle
 \label{eq:chidef}
 \; \; \; .
 \end{equation}
On the other hand, the real part of the dimensionless
spin conductance (\ref{eq:spinconductivity})
is given by
 \begin{equation}
 {\rm Re} [ {G}_s ( \omega ) ]_{\alpha \beta}^{ij}
 = - \pi \delta ( \hbar \omega ) [{D}_s ]_{\alpha \beta}^{ij}
 + [{G}_s^{\prime} ( \omega ) ]_{\alpha \beta}^{ij}
 \label{eq:ReGspin}
 \; \; \; ,
 \end{equation}
where the weight of the $\delta$-function  can be identified with the
dynamic spin stiffness (\ref{eq:Dsdef}), and the paramagnetic contribution is
 \begin{eqnarray}
 [{G}_s^{\prime} ( \omega )]_{\alpha \beta}^{ij}
 & = & \lim_{\bf{q} \rightarrow 0}
 \frac{ {\rm Im} {K}_{\alpha \beta }^{ij}  ( {\bf{q}} , \omega + i 0^{+} )}{\hbar \omega }
 \nonumber
 \\
 & &  \hspace{-5mm} = {\pi}
 \frac{ 1 - e^{- \hbar \omega / T } }{\hbar \omega} \lim_{ {\bf{q}} \rightarrow 0} 
 \frac{1}{N} \sum_{n,m} p_n
 \delta ( E_m - E_n - \hbar \omega )
 \nonumber
 \\
 & & \hspace{10mm} \times  
 \langle n | \hat{J}^{i}_{\alpha} ( {\bf{q}} ) | m \rangle
 \langle m | \hat{J}^{j}_{\beta} ( -  {\bf{q}} ) | n \rangle
 \label{eq:Pspec}
 \; \; \; .
 \end{eqnarray}
The matrix elements of the current operator 
$\hat{\bf{J}}_{\alpha} = [ \hat{J}^x_{\alpha}, \hat{J}^y_{\alpha} 
, \hat{J}^z_{\alpha} ]$
can be related to the matrix elements of the
spin operators $\hat{\bf{S}}_{\bf{q}}$ via the Heisenberg equation 
of motion. Using the fact that the Hubbard interaction  is spin-rotationally
invariant, it is easy so show that to leading order in ${\bf{q}}\cdot {\bf{a}}_{\alpha}$
 \begin{equation}
 i \hbar \frac{ \partial \hat{\bf{S}}_{\bf{q}}}{ \partial t} = [ 
 \hat{\bf{S}}_{\bf{q}} , \hat{T} ]
 = \sum_{\alpha = 1}^d ( {\bf{q}} \cdot {\bf{a}}_{\alpha} ) 
 \hat{\bf{J}}_{\alpha} ( {\bf{q}} )
 \; \; \; .
 \label{eq:EoM}
 \end{equation}
Hence,
 \begin{equation}
 (E_m - E_n ) 
 \langle n | \hat{\bf{S}}_{\bf{q}} | m \rangle = 
  \sum_{\alpha = 1}^d ( {\bf{q}} \cdot {\bf{a}}_{\alpha} ) 
 \langle n | \hat{\bf{J}}_{\alpha} ( {\bf{q}} ) | m \rangle
 \; .
 \end{equation}
Substituting this expression into Eq.(\ref{eq:dynstruc}),
it is easy to show that
 \begin{equation}
 [ {G}_s^{\prime} ( \omega ) ]^{ii}_{\alpha \alpha} 
 =   \frac{ 1 - e^{- \hbar \omega / T}}{\hbar \omega}
 \frac{ \hbar}{2 }
  \lim_{{\bf{q}} \rightarrow 0}
 \frac{ \omega^2}{({\bf{q}} a)^2} {S}^{ii} ( {\bf{q}} , \omega )
 \label{eq:Ptrace}
 \; .
 \end{equation}
Assuming now the
diffusive form (\ref{eq:spindifdyn}) of the dynamic structure factor,
we obtain for the spin diffusion coefficient
 \begin{equation}
 \frac{\hbar {\cal{D}}}{a^2} = \frac{1}{ \chi } \lim_{\omega \rightarrow 0} 
 [{G}_s^{\prime} ( \omega ) ]^{ii}_{\alpha \alpha}
 \label{eq:spindifres}
 \;  .
 \end{equation}
Moreover, it is not difficult to show\cite{Kopietz93} that in the presence
of spin diffusion the dynamic spin stiffness $D_s$ 
vanishes due to a perfect cancellation between the dia- and
paramagnetic contributions in Eq.(\ref{eq:Dsdef}).
Thus, the existence of spin diffusion means that the system
is a normal spin conductor.

Eq.(\ref{eq:spindifres}) can be rewritten in a form which
emphasizes a deep connection between spin diffusion and
charge diffusion 
in disordered electronic systems\cite{Thouless74,Chakravarty91,Kopietz93}.
Defining the rescaled dimensionless spin conductance\cite{footnote1}
 \begin{equation}
 g_s = \left( \frac{L}{a} \right)^{d-2}
 \lim_{\omega \rightarrow 0} 
 [{G}_s^{\prime} ( \omega ) ]^{ii}_{\alpha \alpha}
 \; ,
 \end{equation}
and the energies
 \begin{equation}
 E_{\rm Th} = \frac{ \hbar {\cal{D}}}{L^2}
 \; \; \; , \; \; \; 
 \Delta_s = \frac{1}{N \chi}
 \; ,
 \label{eq:energies}
 \end{equation}
(where $L = a N^{1/d}$ is the linear size of the system)
we obtain from Eq.(\ref{eq:spindifres})
 \begin{equation}
 g_s = \frac{ E_{\rm Th}}{ \Delta_s}
 \label{eq:gsfinal}
 \; .
 \end{equation}
This expression should be compared with the well-known
Thouless formula $g = E_{\rm Th} / \Delta$ for the dimensionless
average conductance of a disordered electronic system.
Here the so-called Thouless energy $E_{\rm Th}$ 
is defined as in Eq.(\ref{eq:energies}) 
(with ${\cal{D}}$ now given by the average charge diffusion coefficient
of the disordered system), and
$\Delta$ is the average level spacing at the Fermi energy.
Thus, Eqs.(\ref{eq:spindifres}) and (\ref{eq:gsfinal}) are nothing
but the  Thouless formula for the spin diffusion problem.
The dimensionless number $g_s$ defined in Eq.(\ref{eq:gsfinal})
is the corresponding Thouless number.
In analogy with disordered electrons, a system with $g_s \gg 1$ 
can be called a {\it{spin metal}}.

\subsection{Spin diffusion in the Hubbard model}

The above analogies are not only interesting from a
formal point of view, but
also useful in practice.
We now show that
Eq.(\ref{eq:spindifres}) offers a new and physically transparent
extrapolation scheme for directly calculating the
spin diffusion coefficient of the Hubbard model.
See Ref.\cite{Kopietz93} for a similar calculation
for the Heisenberg model, and Ref.\cite{Bonca95} for an alternative
method to calculate the spin diffusion coefficient in the
two-dimensional $t-J$-model.
 
After some straightforward manipulations, Eq.(\ref{eq:spindifres})
can be cast into the form
 \begin{equation}
 \frac{ \hbar {\cal{D}}}{a^2} =
 \frac{ t^2}{T \chi} \int_0^{\infty} d s C ( s)
 \; ,
 \label{eq:difres}
 \end{equation}
where the correlation function $C ( s)$ is given by
 \begin{equation}
 C ( s ) = \frac{1}{2 N} \langle
 [ \hat{I} ( s) + \hat{I} ( -s)] \hat{I} \rangle
 \; .
 \label{eq:Csdef}
 \end{equation}
Here $\hat{I} ( s) = e^{i \hat{H} s} \hat{I} e^{- i \hat{H} s }$, and
the (dimensionless)
current operator $\hat{I}$ is 
 \begin{eqnarray}
 \hat{I} & = &
   \frac{1}{2 i} \sum_{\bf{r}}
 \left[ {c}^{\dagger}_{\bf{r}} 
 \frac{\sigma^{z}}{2}
 ( {c}_{{\bf{r}} + {\bf{a}}_{\alpha} }   -
  {c}_{{\bf{r}} - {\bf{a}}_{\alpha} }  ) 
  - \mbox{h.c.}
 \right] 
 \nonumber
 \\
 & = & \sum_{\bf{k}} \sin ( k_x a ) \left[ 
 c^{\dagger}_{ {\bf{k}} \uparrow} c_{ {\bf{k}} \uparrow }
 -
 c^{\dagger}_{ {\bf{k}} \downarrow} c_{ {\bf{k}} \downarrow }
 \right]
 \; ,
 \label{eq:Idef}
 \end{eqnarray}
where $c_{\bf{k}} = N^{-1/2} \sum_{\bf{r}} e^{ -i {\bf{k}} \cdot
  {\bf{r}}} c_{\bf{r}}$.
The bracket in Eq.(\ref{eq:Csdef}) denotes thermal average
with respect to the interacting Hamiltonian $\hat{H} = \hat{T} +
\hat{V}$, see Eqs.(\ref{eq:Tdef}) and (\ref{eq:Vdef}).
Because the kinetic energy operator $\hat{T}$ commutes with the
current operator $\hat{I}$, for $U = 0$ the integral in
Eq.(\ref{eq:difres})
does not exist. 
Then our model is an ideal spin conductor.
This is not surprising, because the diffusive
dynamics in a system without disorder must be a correlation effect.
We would like to emphasize that
Eq.(\ref{eq:difres}) has been derived under the {\it{assumption}} that
the spin dynamics is diffusive. The divergence of the integral for
$U =0$ simply indicates that in this case our assumption is not
correct.

Because for $U =0$ the integral in Eq.(\ref{eq:difres}) is infinite, we
expect that for small $U $ the spin diffusion coefficient diverges
with some power of $t / U$. Of course, for finite $U$ the correlator
$C ( s)$ cannot be calculated exactly, so that we have to make some
physically motivated approximation. A standard approximation
scheme, which has proven to be quite reliable 
for the calculation of the spin diffusion coefficient of the
Heisenberg model at high temperatures\cite{Kopietz93,Moriya56}, 
is the Gaussian
extrapolation of the short-time expansion of $C ( s)$
to long times.
Expanding $C ( s)$ in powers of $s$, 
 \begin{equation}
 C ( s ) = \sum_{n = 0}^{\infty} \frac{ (-1)^n s^{2n}}{ ( 2 n)!}
 C_{2n}
 \; ,
 \label{eq:shorttime}
 \end{equation}
the coefficients $C_{2n}$
can be written in terms of multiple commutators.
Because $C ( -s) = C ( s)$, only even powers of $s$ appear.
The first two coefficients are
 \begin{eqnarray}
 C_0 & = & \langle \hat{I}^2 \rangle
 \label{eq:C0def}
 \; ,
 \\
 C_2 & = & \langle \hat{I} \left[ \left[ \hat{I} , \hat{H} \right] ,
 \hat{H} \right] \rangle
 \nonumber
 \\
 & = &
 \langle \hat{I} \left[ \left[ \hat{I} , \hat{V} \right] ,
 \hat{V} \right] \rangle 
 \label{eq:C2def}
 \; ,
 \end{eqnarray}
where we have used  $[ \hat{I} , \hat{T}] = 0$ and
$\langle \hat{I} \rangle = 0$.
{\it{Assuming}} that the higher coefficients are consistent
with a Gaussian, the long-time extrapolation is
 \begin{equation}
 C ( s ) \approx C_0 \exp \left[ - \frac{C_2 s^2}{2 C_0} \right]
 \label{eq:longtime}
 \; .
 \end{equation}
Then we obtain from Eq.(\ref{eq:difres})
 \begin{equation}
 \frac{ \hbar {\cal{D}}}{a^2} =
 \frac{ t^2}{T \chi} \frac{C_0}{2} \left[ \frac{ 2 \pi C_0}{C_2} \right]^{1/2}
 \; .
 \label{eq:difres2}
 \end{equation}  
Note that so far we have not assumed that the interaction is small, so
that
Eqs.(\ref{eq:C0def})--(\ref{eq:difres2}) are
valid  for arbitrary $U$.
For simplicity, let us now evaluate the coefficients 
$C_0$ and $C_2$ in the limit $ U \rightarrow 0$.
Then the averages in Eqs.(\ref{eq:C0def}) and (\ref{eq:C2def})
are easily calculated with the help of the Wick-theorem.
Specializing to the case of half filling, 
we obtain after a lengthy but straightforward calculation
 \begin{eqnarray}
 C_0  & = & 2 A_d ( T ) 
 \label{eq:C0res}
 \; ,
 \\ 
 C_2  & = & 4 U^2 A_d ( T ) B_d ( T )
 \label{eq:C2res}
 \; ,
 \end{eqnarray}
where
 \begin{eqnarray}
 A_d ( T ) & = &
 \frac{1}{N} \sum_{\bf{k}} \sin^2 ( k_x a) 
 f ( \epsilon_{\bf{k}} / T ) \left[ 1 - f ( \epsilon_{\bf{k}} / T ) 
 \right]
 \; ,
 \label{eq:Addef}
 \\
 B_d ( T ) & = & \frac{1}{N} \sum_{\bf{k}}  
 f ( \epsilon_{\bf{k}} / T ) \left[ 1 - f ( \epsilon_{\bf{k}} / T ) 
 \right]
 \; .
 \label{eq:Bddef}
 \end{eqnarray}
Here $f ( x ) = [ e^x + 1]^{-1}$ is the Fermi function, and the
non-interacting energy dispersion in $d$
dimensions is $\epsilon_{\bf{k}} = 2 d t \gamma_{\bf{k}}$, with
$\gamma_{\bf{k}} = d^{-1} \sum_{\alpha = 1}^{d} \cos ( {\bf{k}} \cdot {\bf{a}}_{\alpha})$.
Away from half filling we should replace $\epsilon_{\bf{k}}
\rightarrow
\epsilon_{\bf{k}} - \mu$, where $\mu$ is the chemical potential.
In the non-interacting limit
$T \chi = \frac{1}{2} B_d ( T)$, so that we finally
obtain from Eq.(\ref{eq:difres2}) in the limit $U \ll t$
 \begin{equation}
 \frac{ \hbar {\cal{D}}}{a^2} =
 2 \sqrt{\pi}
 \frac{  A_d ( T)}{B^{3/2}_d ( T)}
 \frac{ t^2}{ U } 
 \; .
 \label{eq:difres3}
 \end{equation} 
Note that ${\cal{D}}$ diverges for
$U \rightarrow 0$, in agreement with the fact that
without correlations there is no spin diffusion.
Obviously, the spin diffusion coefficient cannot be calculated
by naive perturbation theory in powers of $U$.

In the limit $T \gg t$ we may use
 $A_d ( \infty) = \frac{1}{4}$ and $B_d ( \infty ) =
\frac{1}{8}$,
so that Eq.(\ref{eq:difres3}) reduces to
  \begin{equation}
 \frac{ \hbar {\cal{D}}}{a^2} =
 2 \sqrt{\pi} 
 \frac{ t^2}{ U }
 \; \; , \; \; T = \infty 
 \; ,
 \label{eq:difresTinf}
 \end{equation} 
independent of the dimensionality of the system.
Recall that this result has been derived in the weak coupling limit.
More generally, at $T = \infty$
it is easy to see from Eqs.(\ref{eq:difres}), (\ref{eq:Csdef}), and
(\ref{eq:shorttime}) that
$\hbar {\cal{D}} / a^2$ is proportional to
$t^2 / U$ {\it{for all values of}} $U$.
This follows 
from the fact that the  expansion (\ref{eq:shorttime}) of $C ( s)$ 
is actually an expansion in powers of
$(Us )^2$, because the current operator $\hat{I}$ commutes with
the kinetic energy operator $\hat{T}$. Assuming that the integral
is convergent, we may scale out the $U$-dependence by 
re-defining the integration variable
$s^{\prime} = Us$. This leads trivially to the energy scale $t^2 / U$.
We would like to point out that
at $T = \infty$ the spin diffusion coefficients
of the spin $S = 1/2 $ quantum Heisenberg
antiferromagnet and the half filled Hubbard model at strong coupling
are {\it{not}} identical, although
both are proportional to 
$ t^2/U$ (see Refs.\cite{DeGennes58,Kopietz93}).
The reason is that for $T = \infty$
the value of ${\cal{D}}$ in the Hubbard model is determined
by states with energies larger than $U$, while the mapping to
the Heisenberg model is only valid
for energy scales smaller than $U$.
Only in the interval $t^2 /U \ll T \ll U$
the half filled Hubbard model at strong coupling 
can be expected to have 
the same  spin diffusion 
coefficient as the corresponding Heisenberg
antiferromagnet with exchange coupling $J = 4 t^2 / U$.

Let us now discuss the low-temperature limit $T \ll t$.
Using the fact that for $T \rightarrow 0$
 \begin{equation}
 f ( \epsilon_{\bf{k}} / T ) \left[ 1 - f ( \epsilon_{\bf{k}} / T ) 
 \right] \rightarrow \frac{T}{2dt} \delta ( \gamma_{\bf{k}})
 \; ,
 \label{eq:deltalim}
 \end{equation}
it is  easy to see that Eq.(\ref{eq:difres3}) reduces to
 \begin{equation}
 \frac{ \hbar {\cal{D}}}{a^2} = 2 \sqrt{\pi}
 \frac{  a_d}{b^{3/2}_d}
 \left[ \frac{2 d t}{T} \right]^{1/2}
 \frac{ t^2}{ U }
 \; \; \; , \; \; \; T \ll t  
 \; ,
 \label{eq:difresT}
 \end{equation} 
where the numerical constants $a_d$ and $b_d$ are
 \begin{eqnarray}
 a_d   & = &
 \frac{1}{N} \sum_{\bf{k}} \sin^2 ( k_x a)
 \delta ( \gamma_{\bf{k}})
 \; ,
 \label{eq:addef}
 \\ 
 b_d  &  = &
 \frac{1}{N} \sum_{\bf{k}} 
 \delta ( \gamma_{\bf{k}})
 \; \; \; , \; \; \; d \neq 2
 \; .
 \label{eq:bddef}
 \end{eqnarray}
In $d=2$ the integral in Eq.(\ref{eq:bddef}) is
logarithmically divergent (for $N \rightarrow \infty$),
so that at low temperatures $b_2$ is given by
 \begin{equation}
 b_2 \approx \frac{4}{\pi^2} \ln ( 4 t / T)
 \; .
 \end{equation}
Because  we have assumed that the
system is in the paramagnetic state, Eq.(\ref{eq:difresT})
should be valid for temperatures above the magnetic ordering
temperature $T_N$. Keeping in mind that in $d \leq 2$ 
there is no long range order at any finite temperature, and that
in $d > 2$ the ordering temperature $T_N$ is exponentially
small at weak coupling, Eq.(\ref{eq:difresT}) 
describes the low-temperature behavior
of the spin diffusion coefficient in a wide range of temperatures
that are small compared with the band-width $4 d t$. 
Although the precise numerical value of the  prefactor 
in Eqs.(\ref{eq:difres3}) and (\ref{eq:difresT})  
depends on our Gaussian extrapolation scheme,
the energy scale $t^2 / U$ in Eq.(\ref{eq:difres3}) and the 
low-temperature behavior given in Eq.(\ref{eq:difresT})
should be independent of the
details of the extrapolation method.

\section{Summary and Conclusions}

In this work we have used
analogies between charge and spin response functions 
of itinerant magnets to 
clarify the concept of the static and dynamic
spin stiffness.
Starting from the general Kubo formula for the
relevant linear response functions, we have
shown that the {\it{dynamic}} 
spin stiffness $D_s$ is the precise spin analog
of the charge  stiffness $D_c$.
The usual (static) spin stiffness $\rho_s$ is the spin analog of the
superfluid stiffness $\rho_c$, and is only finite
in the presence of long-range magnetic order.
Considering the fact that the charge stiffness 
has recently received a lot of attention\cite{Scalapino92,Castella95,Zotos96},
it is rather surprising that the corresponding
quantity $D_s$ has not been studied.
Partially, this might be due to the fact that
$D_s$ has often been confused with the
static spin stiffness $\rho_s$.
Table \ref{tab:analogy} summarizes 
analogous quantities.
We would like to emphasize again that a finite value of $D_s$ 
does not
imply the existence of long-range
magnetic order. A simple example is the
Hubbard model for $U = 0$, where $D_s > 0$ but
$\rho_s = 0$. A value $D_s > 0$ simply means that
the system is an ideal spin conductor, so that
the spin transport is not diffusive.
The analogy with charge transport is obvious: 
an ideal conductor has a finite charge stiffness $D_c > 0$, implying
an infinite conductivity.
But a perfect conductor is not necessarily a superconductor.
Only in the latter case $\rho_c > 0$.

The low-frequency behavior of the dynamic spin conductance
$G_s ( \omega)$ can be used to classify 
the spin dynamics into three categories:
ideal spin conductors, normal spin conductors, and spin insulators.
In Sec.\ref{sec:diffusion} we have further analyzed 
a particular class of normal spin conductors, 
namely systems with diffusive spin dynamics.
In this case
the dynamic spin stiffness vanishes, but the spin conductance
$G_s ( \omega )$ has a finite limit for vanishing frequency,
which is proportional to the
spin diffusion coefficient.
The concept of the spin conductance  and the associated
Thouless number offers a new  and physically transparent
extrapolation scheme for calculating the spin diffusion
coefficient.

We hope that our work  will
stimulate further research in this field.
Numerical calculations of $G_s ( \omega)$
and $D_s$ in strongly correlated itinerant magnets
are called for. In particular, 
by varying some external parameter
(such as temperature, doping, or interaction strength),
it might be possible to observe
transitions between the three types of
spin transport discussed above. 
Numerical calculations of the temperature-dependence
of the spin conductance might also give evidence for 
spin-charge separation in strongly
correlated systems:
Very recently  Si\cite{Si97} pointed out that
the existence of spin-charge separation
manifests itself
in different temperature-dependencies of the
spin- and charge conductances.

\section*{Acknowledgments} 
I would like to thank H. G. Evertz for a discussion
during a workshop on the
{\it{Role of Dimensionality in Correlated Electronic Systems}} at
Villa Gualino (Torino, Italy), which eventually
motivated me
to write this paper. 
This work was partially supported
by a Heisenberg Fellowship of the Deutsche Forschungsgemeinschaft, and
by the ISI Foundation and EU HC\&M Network ERBCHRX-CT920020.
 
\widetext

\newpage

\squeezetable
\begin{table}
\caption{
Analogous quantities characterizing charge- and spin dynamics.}
\begin{tabular}{llp{5cm}}
Charge &Spin&Physical meaning \\
\tableline
& & \\
charge response function  $K_{\alpha \beta} ( {\bf{q}} , \omega )$
&
spin response function ${K}_{\alpha \beta}^{ij} ( {\bf{q}} , \omega )$
&
linear current response to an external vector potential
\\
conductance $G_{\alpha \beta} ( \omega )$ & spin conductance
$[{G}_s ( \omega )]_{\alpha \beta}^{ij}$ & 
response to a time-dependent,  
spatially constant vector potential 
\\
${\bf{q}}$-dependent superfluid stiffness\tablenote{In the book by
  Tinkham\cite{Tinkham75} the transverse eigenvalue
of this tensor is denoted by $K ( {\bf{q}})$.}
$[ \rho_c 
( {\bf{q}}) ]_{\alpha \beta }$ &
${\bf{q}}$-dependent spin stiffness
$[{\rho}_s ( {\bf{q}}) ]^{ij}_{\alpha \beta }$ &
response to a time-independent,
spatially varying vector potential
\\
superfluid stiffness $\rho_c$ &
spin stiffness ${\rho}_s$ &
probes long-range correlations
(superconducting or magnetic)
\\
charge stiffness\tablenote{Also known as Drude weight} $D_c$ &
dynamic spin stiffness ${D}_s $
&
finite for ideal (charge or spin) conductor
\end{tabular}
\label{tab:analogy}
\end{table}


\begin{thebibliography}{99}
\bibitem{footnote1}
For a system of linear size $L$ the conductivity $\sigma$ is related to the
usual dimensionless conductance $g$ via
$\sigma = (e^2 / h ) L^{2-d} g$.
Thus, our function $G_{\alpha \beta }$ 
in Eq.(\ref{eq:conductance}) is related to the dimensionless
conductance $g$ via $G = (L / a )^{2-d} g$.
\end{thebibliography}
\end{document}